\documentclass[aps,pra,twocolumn,groupedaddress,showpacs,amsfonts,amssymb,superscriptaddress]{revtex4}

\usepackage{graphics}
\usepackage{epsfig}
\usepackage{amsmath}

\begin{document}


\title{Bounds on classical information capacities for a class of quantum memory channels}


\author{Garry Bowen}
\email{gab30@damtp.cam.ac.uk}
\affiliation{Centre for Quantum Computation, DAMTP, 
University of Cambridge, Cambridge CB3 0WA, UK}
\affiliation{Department of Computer Science, University of Warwick, Coventry CV4 7AL, UK}
\author{Igor Devetak}
\email{devetak@us.ibm.com}
\affiliation{IBM T.~J.~Watson Research Center, Yorktown Heights, NY 10598, USA}
\author{Stefano Mancini}
\email{stefano.mancini@unicam.it}
\affiliation{Dipartimento di Fisica, Universit\`{a} di Camerino, 
I-62032 Camerino, Italy}
\date{\today}

\begin{abstract}
The maximum rates for information transmission through noisy quantum channels has primarily been developed for memoryless channels, where the noise on each transmitted state is treated as independent.  Many real world communication channels experience noise which is modelled better by errors that are correlated between separate channel uses.  In this paper, upper bounds on the classical information capacities of a class of quantum memory channels are derived.  The class of channels consists of indecomposable quantum memory channels, a generalization of classical indecomposable finite--state channels.
\end{abstract}


\maketitle

\section{Introduction}

Communication of information requires an encoding of the information into a physical system.  The laws of physics therefore govern the limits on processing and communication of information.  By modelling real world noise in terms of simpler models, the maximum rate for information transfer may be obtained.  The seminal work of Shannon \cite{shannon48a} showed that a memoryless noisy channel, where the noise acts independently on each symbol sent through the channel, can be parameterized by a single quantity, the capacity of the channel $C$.  Shannon defined the capacity as the maximum rate that information may be sent through a channel, and he also showed that there exist codes that asymptotically achieve this rate with a vanishing probability of error.  The work of Shannon has been extended to include channels with ``memory'', where the noise is no longer independent of past channel uses \cite{gallager}, or where the channel consists of an arbitrary transformation of the input states \cite{verdu94}.

Encoding classical information into quantum states of physical systems gives a physical implementation of the constructs of information theory.  The majority of research into quantum communication channels has focused on the memoryless case, although there have been a number of important results obtained for quantum channels with correlated noise operators, or more general quantum channels \cite{macchiavello02,hamada02,daffer02,hayashi03}.

Recently a model for quantum channels with memory has been proposed that can consistently define quantum channels with Markovian correlated noise \cite{bowen03}.  The model also extends to describe channels that act on transmitted states in such a way that there is no requirement for interaction with an environment within the model.

In this paper a single upper bound on the capacity of arbitrary indecomposable quantum channels is derived.

\section{Unitary Representations of Quantum Channels}

A quantum channel is defined as a completely positive, trace preserving map from the set of density operators to itself.  Any such map may be represented as a unitary operation between the system state and an environment with a known initial state \cite{kraus}.

\subsection{Unitary Representation of Memoryless Channels}

Memoryless quantum channels act on each input state independently of the previous input or output states.
For a single channel use the output state is given by,
\begin{equation}
\Lambda\rho_Q = \mathrm{Tr}_E \Big[ U_{QE} \big( \rho_Q \otimes |0_E\rangle \langle 0_E| \big) U_{QE}^{\dag} \Big]
\end{equation}
with $\rho_Q$ the input state, $|0_E\rangle\langle 0_E|$ the initial state of the environment, $U_{QE}$ a unitary operation between the state $Q$ and environment $E$, and $\Lambda\rho_Q$ the output state.
For a sequence of transmissions through a memoryless channel, the output state is given by,
\begin{align}
\Lambda^{(n)}\rho_Q &= \mathrm{Tr}_{E} \Big[ 
U_{n,E_n}...U_{1,E_1} \big(\rho_Q \otimes |0_{E_1} ... 0_{E_n} \rangle \langle 0_{E_1} ... 0_{E_n} | \big) \nonumber \\
&\phantom{=}\: \times U_{1,E_1}^{\dag} ...  U_{n,E_n}^{\dag}\Big] \nonumber \\
&= \big(\Lambda_{n} \otimes ... \otimes \Lambda_{1} \big) \rho_Q
\label{eqn:memoryless_model}
\end{align}
where the state $\rho_Q$ now represents a (possibly entangled) input state across the $n$ channel uses, the unitary operations $U_{i,E_i}$ are all identical, and the environment state is a product state $|0_{E_1} ... 0_{E_n} \rangle = |0_{E_1} \rangle \otimes ... \otimes |0_{E_n} \rangle$.

\subsection{A Unitary Model for Memory Channels}

\begin{figure}
\centering
\includegraphics[width=2.5in]{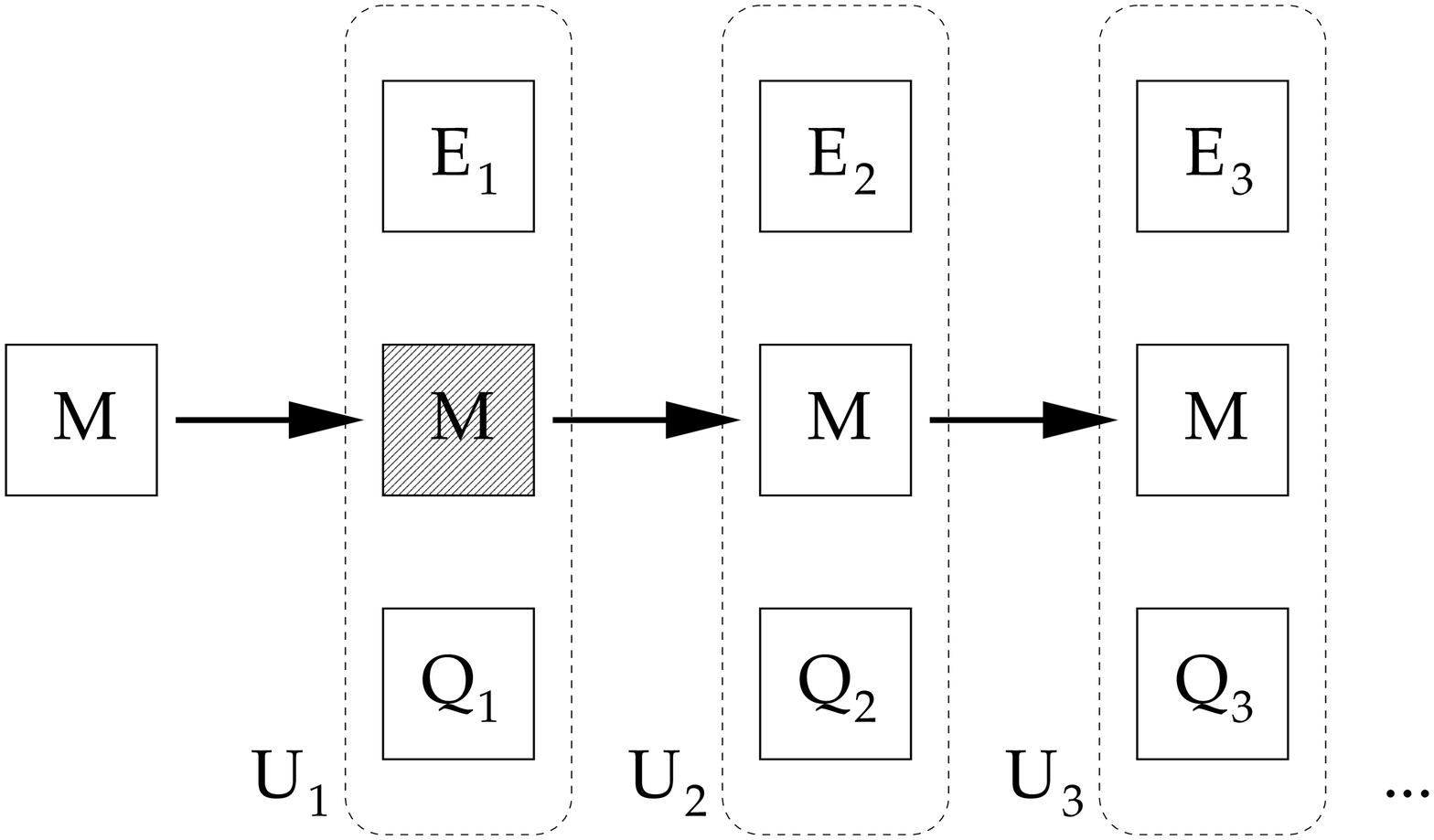}
\caption{Diagram of the model for a quantum memory channel.  The initial memory state interacts with each transmitted quantum state $Q_i$ and environment $E_i$.  The correlations between the error operators on each state $Q_i$ are determined by the unitary operation $U_i$ and the memory state at each stage of the channel evolution.}
\label{fig:memory_channel}
\end{figure}

One model of a quantum memory channel is where each state going 
through the channel acts with a unitary interaction \textit{on the same 
channel memory state}, as well as an independent environment.  The 
backaction of the channel state on the message state therefore gives 
a memory to the channel.
The general model thus includes a channel memory $M$, and the independent 
environments for each qubit $E_i$.  Hence,
\begin{align}
\Lambda^{(n)}\rho_Q &= \mathrm{Tr}_{ME} \Big[ 
U_{n,ME_n}...U_{1,ME_1} \big(\rho_Q \otimes \omega_M \nonumber \\
&\phantom{=}\: \otimes |0_{E_1} ... 0_{E_n} \rangle \langle 0_{E_1} ... 0_{E_n} | \big) U_{1,ME_1}^{\dag} ...  U_{n,ME_n}^{\dag}\Big] \nonumber \\
&= \mathrm{Tr}_{M} \Big[ \Lambda_{n,M} ... \Lambda_{1,M} 
\big(\rho_Q \otimes \omega_M \big) \Big]
\label{eqn:memory_model}
\end{align}
where $\omega_M$ is the initial memory state, $\rho_Q$ and $\Lambda^{(n)}\rho_Q$ are the
input and the output states of the channel, respectively, and
the trace over the environment is over all environment states.
Figure \ref{fig:memory_channel} illustrates the action of the unitary operators on the input, memory and environment states representing the channel.

If the unitaries factor into independent unitaries acting on the 
memory and the combined state and environment, that is, $U_{n,ME_n} = 
U_{n,E_n}U_M$, then the memory traces out and we have a memoryless 
channel.  If the unitaries reduce to $U_{n,M}$, we can call it a 
\textit{perfect} memory channel, as no information is lost to the 
environment.
The mapping of the memory state under the unitary operation also corresponds to a quantum channel on the memory state.  Memory channel representations where the action on the memory state is independent of the input state are termed \textit{fixed--point} memory channels, as the memory state will have a fixed point under the action of the represetation $\Phi[\rho_Q]\omega_M = \omega_M$ for all $\rho_Q$.  Fixed--point channels may also be seen to be symbol independent (SI), as the previous input states do not affect the action of the channel on the current input state.  This is opposed to channels with intersymbol interference (ISI), where the previous input state affects the action of the channel on the current input.  An extreme example of an ISI channel is the quantum shift channel, where each input state is replaced by the previous input state.
There exist examples of SI channels that are not fixed point channels.  The ``classical'' memory channel defined in terms of unitary Kraus error operators by,
\begin{align}
U_{Q_iME_i}|\phi_{Q_i}\rangle |j_M\rangle |0_{E_i}\rangle &= \sum_k 
\sqrt{p_{k|j}} V_k^{(i)}|\phi_{Q_i}\rangle |k_M\rangle 
|j_{E_i}\rangle \nonumber \\
\label{eqn:markovmem} 
\end{align}
for $p_{k|j}$ the transition probabilities of a Markov chain, is not a fixed point channel, but displays no ISI.  In this case, however, the channel may equivalently be defined by,
\begin{align}
\tilde{U}_{Q_iME_i}|\phi_{Q_i}\rangle |j_M\rangle |0_{E_i}\rangle &= \sum_k 
\sqrt{p_{k|j}} V_k^{(i)}|\phi_{Q_i}\rangle |k_M\rangle 
|j,k_{E_i}\rangle \nonumber \\
\label{eqn:markovmem2} 
\end{align}
where the environment states $|j,k_{E_i}\rangle$ are orthogonal.  This results in a fixed--point channel with the equivalent output as obtained from (\ref{eqn:markovmem}).  If the representations of a given memory channel are treated as an equivalence class under the input--output action of the channel, then for every channel of the above type, there exists a fixed--point representation of the channel.  Whether this is true for arbitrary ISI channels is not known.
We may also conjecture that the types of memory channels that display \textit{only} ISI may be represented as perfect memory channels.

\section{Entropic Upper Bounds on the Capacity}

The Fano inequality, combined with the Holevo upper bound on the accessible information, provides an entropic upper bound on the classical information capacity of any quantum channel.  Utilizing classical--quantum states of the form,
\begin{equation}
\rho_{RQ} = \sum_i p_i |r_i\rangle \langle r_i| \otimes \rho_Q^i
\label{eqn:class_quant1}
\end{equation}
where the $|r_i\rangle$ form an orthonormal set.  It is possible to derive the Holevo bound from the von Neumann mutual information bound over these classical--quantum states.  To view the upper bound in a more physically motivated setting we can note that any separable state $\rho_{RQ}$ can be extended, in a larger Hilbert space, to a state $\rho_{R\bar{R}Q}$ in the form of (\ref{eqn:class_quant1}).  To show this note that any separable state may be written in the form,
\begin{equation}
\rho_{RQ} = \sum_j p_j \rho_R^j \otimes \rho_Q^j
\end{equation}
and each $\rho_R^j$ may be purified into a direct sum of Hilbert spaces $\mathcal{H}_{\bar{R}}^j$ with orthogonal support, such that,
\begin{equation}
\rho_{R\bar{R}Q} = \sum_j p_j |\bar{r}^j_{R\bar{R}_j}\rangle \langle \bar{r}^j_{R\bar{R}_j}| \otimes \rho_Q^j
\label{eqn:class_quant2}
\end{equation}
where $\mathrm{Tr}_{\bar{R}} \big[ |\bar{r}^j_{R\bar{R}_j}\rangle \langle \bar{r}^j_{R\bar{R}_j}| \big] = \rho^j_R$.  Due to the monotonicity of the von Neumann mutual information $S(R\bar{R} : Q) \geq S(R:Q)$, the upper bound may be expressed as,
\begin{equation}
C \leq \max_{\rho_{RQ} \in \mathcal{D}} S(R:\Lambda Q)
\end{equation}
where $\mathcal{D}$ is the set of all separable states.  The maximum amount of mutual information that may be generated through a quantum channel $\Lambda$ is therefore bounded by the maximum amount of ``classical'' correlation that may be shared by states through the channel.
The upper bound on the classical capacity of a finite memory channel, which includes the classically correlated channels, is given by \cite{bowen03},
\begin{equation}
C = \liminf_{n\rightarrow \infty}\max_{\rho_{RQ} \in \mathcal{D}} \frac{1}{n} S(R:\Lambda^{(n)}[\omega_M]Q)
\label{eqn:mem_holevo_upperbound}
\end{equation}
where $\Lambda^{(n)}[\omega_M]Q = \Lambda^{(n)}[\omega_M]\rho_Q$ is the action of the channel on the input state, with initial memory state $\omega_M$, and $\mathcal{D}$ the set of separable states.
This bound has been shown to be attainable whenever the channel is described by (\ref{eqn:markovmem2}) with unitary Kraus operators $A_{i_k}$, and the initial error probabilities are equal to the steady state probabilities for the regular Markov chain \cite{bowen03}.

\section{Indecomposable Channels}

An indecomposable channel is one where the long-term behavior of the channel is independent of the initial channel state.  Memory channels with Markov correlated noise are an example of indecomposable channels.

\subsection{Trace Distance and Indecomposable Memory Channels}
We begin by defining the \textit{trace distance} of both probability distributions and density operators.  For probability distributions the trace distance is defined as $\| P - Q \| = \frac{1}{2}\sum_i |p_i - q_i|$,
for distributions $P = \{ p_i \}$ and $Q = \{ q_i \}$, where both distributions share the index set $i \in \mathcal{I}$ \cite{nielsen}.  For density operators the trace distance is defined as \cite{nielsen},
\begin{equation}
\| \,\rho - \omega \,\| = \frac{1}{2} \mathrm{Tr} |\rho - \omega |
\end{equation}
where $| F | = \sqrt{F^{\dag}F}$, taking the positive square root.
A finite-memory quantum channel is \textit{indecomposable} if for any input state $\rho$ and $\epsilon > 0$ there exists an $N(\epsilon)$ such that for $n \geq N(\epsilon)$,
\begin{equation}
\| \, \omega_M(n,\rho) - \sigma_M(n, \rho) \| \leq \epsilon
\label{eqn:indecomp}
\end{equation}
where $\omega_M(n,\rho)$ and $\sigma_M(n,\rho)$ are the memory states after $n$ uses of the channel for the intial memory states $\omega_M$ and $\sigma_M$ respectively.  The long term behavior of an indecomposable channel is therefore independent of the initial memory state.  Fixed point channels for which the map on the memory state is a strictly contractive mapping $\| \Phi\omega_M - \Phi\sigma_M\| < \| \omega_M - \sigma_M \|$, are automatically indecomposable.
The behavior of the memory state is of fundamental importance due to the fact that memory channels that may be represented in the form of (\ref{eqn:memory_model}) are \textit{memory continuous}.  A channel is memory continuous if for any $\epsilon > 0$ there exist a $\delta >0$, such that,
\begin{align}
\|\,\omega_M - \sigma_M\| \leq \delta \implies \|\Lambda[\omega_M]\rho_Q - \Lambda[\sigma_M]\rho_Q\| \leq \epsilon .
\label{eqn:mem_cont}
\end{align}
Note that the trace distance is monotonic $\| \rho - \omega \| \geq \|\Psi\rho - \Psi\omega\|$ for any trace preserving quantum operation $\Psi$.  Hence, for $\Psi$ the partial trace operation $\|\rho_{RQ} - \omega_{RQ}\| \geq \|\rho_Q - \omega_Q\|$.  Furthermore, due to the unitary invariance of the trace distance, we find,
\begin{align}
&\| \Lambda [\omega_M] \rho_Q - \Lambda [\sigma_M]\rho_Q \| \nonumber \\
&\leq \big\|U_{QME} \big(\rho_Q \otimes \omega_M \otimes |0_{E} \rangle \langle 0_{E}| \nonumber \\
&\phantom{=}\qquad\qquad\qquad\:- \rho_Q \otimes \sigma_M \otimes |0_E\rangle \langle 0_E| \big) U_{QME}^{\dag} \big\| \\
&= \big\|\,\rho_Q \otimes \omega_M \otimes |0_{E} \rangle \langle 0_{E}| - \rho_Q \otimes \sigma_M \otimes |0_E\rangle \langle 0_E| \big\| \\
&= \| \,\omega_M - \sigma_M \|
\end{align}
and (\ref{eqn:mem_cont}) is satisfied for all memory channels of the form of (\ref{eqn:memory_model}), by simply making $\delta = \epsilon$.

\subsection{A Single Upper Bound on the Capacity}

Following Gallager's derivation for classical finite state channels (FSC) \cite{gallager}, two classical capacities may be defined.  The lower capacity $\underline{C}$ and the upper capacity $\overline{C}$ are defined as,
\begin{align}
\underline{C} &= \lim_{n\rightarrow \infty} \frac{1}{n} \min_{\omega_M} \max_{\rho_{RQ}\in \mathcal{D}} S\Big( R:\Lambda^{(n)}[\omega_M] Q \Big) \nonumber \\
\overline{C} &= \lim_{n\rightarrow \infty} \frac{1}{n} \max_{\omega_M} \max_{\rho_{RQ}\in \mathcal{D}} S\Big( R:\Lambda^{(n)}[\omega_M] Q \Big) \; .
\end{align}
It is obvious from the definitions that $\underline{C} \leq \overline{C}$, and we wish to determine channels for which equality holds.
In order to bound the difference in entropies for the output states of channels with different initial memory states, we utilize Fannes inequality \cite{fannes73},
\begin{equation}
\big|S(\omega) - S(\sigma)\big| \leq \| \, \omega - \sigma \| \log d + \frac{\log e}{e} \; ,
\end{equation}
where $d$ is the dimension of the Hilbert space for the states $\omega$ and $\sigma$.
From this we can see that,
\begin{align}
\frac{1}{n}\Big| &S\big(\Lambda[\omega]\rho\big) - S\big(\Lambda[\sigma] \rho\big) \Big| \nonumber \\
&\leq \frac{1}{n}\bigg[ \big\| \Lambda[\omega]\rho - \Lambda[\sigma] \rho \big\| \log d^{\, n} + \frac{\log e}{e} \bigg] \\
&= \big\| \Lambda[\omega]\rho - \Lambda[\sigma] \rho \big\| \log d + \frac{\log e}{ne}
\end{align}
and by showing the trace distance of the output states may be made arbitrarily small for any input state $\rho$, the entropies of the output states must converge asymptotically.
Given any $\epsilon > 0$ then there exists an $N(\epsilon)$ such that the trace distance between the memory states is less than $\epsilon$.  Thus for $n \gg N(\epsilon)$ we have,
\begin{align}
\frac{1}{n}\Big| &S\big(\Lambda[\omega]\rho\big) - S\big(\Lambda[\sigma] \rho\big) \Big| \nonumber \\
&\leq \frac{1}{n}\bigg[ \Big| S\big(\Lambda[\omega(N)]\rho\big) - S\big(\Lambda[\sigma(N)] \rho\big) \Big| + N(\epsilon) \log d \bigg] \nonumber \\
&\leq \frac{1}{n}\bigg[ (n-N(\epsilon))\big\|\Lambda[\omega(N)]\rho - \Lambda[\sigma(N)] \rho\, \big\| \log d \nonumber \\
&\phantom{=}\qquad + \frac{\log e}{e} + N(\epsilon) \log d \bigg] \nonumber \\
&\leq \epsilon \log d + \frac{\log e}{ne} + \frac{N(\epsilon) \log d}{n}(1-\epsilon)
\label{eqn:ave_ent_bound}
\end{align}
where $\Lambda[\omega(N)]$ and $\Lambda[\sigma(N)]$ denote the channels with initial memories $\omega_M$ and $\sigma_M$ acting on input states $N(\epsilon)+1$ to $n$, and the maximum difference in entropy for the first $N(\epsilon)$ output states is bounded by $N(\epsilon) \log d$.  Therefore, taking $n\rightarrow \infty$ in (\ref{eqn:ave_ent_bound}), the average entropies converge asymptotically for any state $\rho$.  The difference between the upper bounds $\overline{C}$ and $\underline{C}$ must therefore also converge asymptotically.

\section{Conclusion}

A single upper bound on the capacity for the class of indecomposable quantum memory channels has been derived.

\end{document}